\documentstyle[12pt,epsfig]{article}
\def\beq{\begin{equation}}
\def\eeq{\end{equation}}
\def\bea{\begin{eqnarray}}  \def\eea{\end{eqnarray}}
\def\lsim{\raise0.3ex\hbox{$<$\kern-0.75em\raise-1.1ex\hbox{$\sim$}}}
\def\gsim{\raise0.3ex\hbox{$>$\kern-0.75em\raise-1.1ex\hbox{$\sim$}}}
\def\1{{\rm 1\mskip-4.5mu l} }
\newcommand{\noi}{\noindent}
\renewcommand{\thefootnote}{\fnsymbol{footnote}}
\begin{document}

\begin{center}
\vspace*{1 truecm}
{\Large \bf Non-saturation of the J/$\psi$ suppression at large } \\
{\Large \bf transverse energy in the comovers approach} \\[8mm]

{\bf A. Capella and E. G. Ferreiro}\par
{\it Laboratoire de Physique Th\'eorique}\footnote{Unit\'e Mixte de Recherche - CNRS - 
UMR N$^{\circ}$ 8627} \\ {\it Universit\'e
de Paris-Sud, B\^atiment 210, F-91405 Orsay Cedex, France}\\
[5mm]
{\bf A. B. Kaidalov}\par
{\it ITEP, B. Cheremushkinskaya, 25, 117259 Moscow, Russia}
\end{center}

\vskip 1 truecm
\begin{abstract}
We show that, contrary to recent claims, the $J/\psi$ suppression resulting from its interaction
with comovers does not saturate at large transverse energy $E_T$. On the contrary, it shows a
characteristic structure - change of curvature near the knee of the $E_T$ distribution - which is
due to the $E_T$ (or multiplicity) fluctuation, and agrees with recent experimental results.
\end{abstract}

\vskip 4 truecm

\noindent LPTHE Orsay 00-21  \par
\noindent February 2000

\newpage
\pagestyle{plain}
An interesting result of the last run (1998 data) of the NA50 collaboration at CERN on the
transverse energy ($E_T$) dependence of $J/\psi$ suppression in $PbPb$ collisions, is the
observation \cite{1r} of a convexity at large $E_T$. More precisely, for $E_T \ \gsim$~100 GeV
(which corresponds to the so-called knee of the $E_T$ distribution \cite{1r}), the slope of the
ratio $R(E_T)$ of $J/\psi$ over Drell-Yan (DY) cross-sections increases with increasing $E_T$.
In sharp contrast with this result, models of the $J/\psi$ suppression in non-quark-gluon plasma
(QGP) scenarios [2--9]
- such as the one based on the interaction of the $J/\psi$ with
comovers - exhibit a clear saturation at large $E_T$. \par

In this work, we show that the above feature of the comovers model is only true up to the knee
of the $E_T$ distribution ($E_T \sim$ 100 GeV). Beyond this value, we enter into the tail of
the $E_T$ distribution - where the increase in $E_T$ is due to fluctuations. This fluctuation,
which has not been taken into account in most calculations, produces a corresponding increase
in the density of comovers - which, in turn, increases the $J/\psi$ suppression at large $E_T$.
\par

In order to illustrate this phenomenon we use the model introduced in ref. \cite{2r}. Here, as
in most non-QGP models, the $J/\psi$ suppression is due to two mechanisms~: absorption of the
pre-resonant $c\bar{c}$ pair with nucleons (the so-called nuclear absorption) and the
interaction of the $J/\psi$ with comovers. The corresponding $J/\psi$ survival probabilities
are given by \cite{2r}.  

\beq 
S^{abs}(b, s) = {\left \{ 1 - \exp [- A \ T_A(s) \ \sigma_{abs}] \right \} \left \{ 1 - \exp
\left [- B\ T_B(b-s) \ \sigma_{abs}\right ]\right \} \over \sigma_{abs}^2 \ AB \ T_A(s) \
T_B(b-s)} \ , \label{1e} \eeq

\beq S^{co}(b, s) = \exp \left \{ - \sigma_{co} \ N_y^{co}(b,
s)\   \ln \left ( {N_y^{co}(b, s) \over N_f} \right )\   \right \} \ .
 \label{2e} \eeq

\par

\noindent
The survival probability $S^{co}$ depends on the density of comovers $N_y^{co}(b,s)$ in the
rapidity region of the dimuon trigger $2.9 < y_{lab} < 3.9$
and $N_f = 1.15$~fm$^{-2}$ \cite{2r,4r} is the corresponding
density in $pp$ collisions.
In order to compute $N_y^{co}$, various
hadronic models have been used in the literature. For instance in ref. \cite{4r} it has been
assumed that the hadronic multiplicity is proportional to the number of participant nucleons
(the so-called wounded nucleon model), while in ref. \cite{2r} a formula based on the dual
parton model (DPM) was used - which includes an extra term proportional to the number of binary
interactions. In this paper we use the DPM formula (eq. (6) of \cite{2r}). In both cases, the
calculations do not include the fluctuations mentioned above and, therefore, cannot be applied
beyond the knee of the $E_T$ distribution - where the increase in $E_T$ (or
multiplicity) is entirely due to fluctuations. In order to introduce these fluctuations, it is
convenient to recall the other formulae needed to calculate the $J/\psi$ suppression. \par

At fixed impact parameter $b$, the $J/\psi$ cross-section is given by \cite{2r}   

 \beq 
\label{3e} 
\sigma_{AB}^{\psi}(b) = {\sigma_{pp}^{\psi} \over
\sigma_{pp}} \int d^2s \  m(b,s) \ S^{abs}(b,s) \ S^{co}(b, s)\  , \eeq

\noi where $m(b, s) = AB \  \sigma_{pp} \ T_A(s) \ T_B(b - s)$. The corresponding one for DY pair
production is obtained from (\ref{3e}) putting $\sigma_{abs} = \sigma_{co} = 0$ (i.e. $S^{abs}
= S^{co} = 1)$ and  is proportional to $AB$. In this way we can compute the ratio of $J/\psi$ over
DY as a function of the impact parameter. Experimentally, however, the ratio $R(E_T)$, is given
as a function of the transverse energy $E_T$ measured by a calorimeter, in the rapidity interval
$1.1 < y_{lab} < 2.3$. In order to compute $R(E_T)$ we have to know the correlation $P(E_T, b)$
between $E_T$ and impact parameter, which is given by \cite{2r}

\beq
\label{4e}
P(E_T , b) = {1 \over \sqrt{2 \pi \ q\  a\ E_T^{NF}(b)}} \exp \left [ - { E_T - E_T^{NF}(b)
\over 2q \ a\ E_T^{NF}(b)} \right ]^2\ .
\eeq

\noi Here

\beq
\label{5e}
E_T^{NF}(b) = q\ N_{cal}^{co}(b) + k [ A - m_A(b)] E_{in} \ , 
\eeq

\noi $m_A(b)$ is the number of participants of $A$ (at fixed impact parameter), $E_{in} =
158$~GeV/c is the beam energy and $k = 1/4000$ \cite{2r}. In (\ref{4e}) and (\ref{5e})
$N_{cal}^{co}(b)$ is obtained by integrating the comover density $N_y^{co}(b,s)$ over $d^2s$, and
$dy$ (in the rapidity range of the $E_T$ calorimeter). The second term in (\ref{5e}) was
introduced in ref. \cite{2r} in order to reproduce the correlation between $E_T$ and the energy
$E_{ZDC}$ of the zero degree calorimeter. It was interpreted as due to intra-nuclear cascade -
which is present here due to the location in rapidity of the NA50 calorimeter. This term is
sizable for peripheral collisions, when many spectator nucleons are present, and dies away for
central ones. The parameters $q=0.56$ and $a=0.94$ are obtained from a fit to the minimum bias $E_T$
distribution at large $E_T$. The parameter $q$ gives the relation between multiplicity of comovers
(positive, negative and neutrals) and the $E_T$ of the NA50 calorimeter (which contains only
neutrals). The product $qa$ controls the width of the $E_T$ distribution at fixed $b \sim 0$. The
$J/\psi$ and DY cross-section at fixed $E_T$ are then given by

\beq
\label{6e}
{d\sigma^{\psi (DY)} \over dE_T} = \int d^2b \ \sigma_{AB}^{\psi (DY)} \ P(E_T, b)\ .
\eeq

The quantity $E_T^{NF}(b)$ in eq. (\ref{5e}) does not contain fluctuations - hence the index
$NF$. This is obvious from the fact that the parameter $a$ is not present in (\ref{5e}). In order
to see it in a more explicit way, we plot in Fig.~1 the quantity

\beq
\label{7e}
F(E_T) = E_T/E_T^{NF}(E_T) \ ,
\eeq

\noi where

\beq \label{8e}
E_T^{NF} (E_T) = {\int d^2b \ E_T^{NF}(b) \ P(E_T,b) \over \int d^2b \ P(E_T, b)}\ . \eeq

\noi We see that $E_T^{NF}$ coincides with $E_T$ only up to the knee of the $E_T$ distribution.
Beyond it, $E_T^{NF}$ is smaller than the true value of $E_T$. This difference is precisely due
to the $E_T$ fluctuation. \par

As discussed above, in order to compute the ratio $R(E_T)$ beyond the knee of the $E_T$
distribution it is necessary to introduce in $N_y^{co}$ the $E_T$ (or multiplicity) fluctuations
responsible for the tail of the distribution. 
In order to do so, we use the experimental observation that multiplity
and $E_T$ distributions have approximately the same shape. This indicates
that the fluctuations in $E_T$ are mainly due to fluctuations in
multiplicity - rather than in $p_T$. This leads to the following replacement
in eq. (\ref{2e}): 

\beq \label{9e} 
N_y^{co}(b,s) \to N_y^{co}(b,s)\ F(E_T) \ . \eeq 

\renewcommand{\thefootnote}{{1}}

\noi In this way the results for the ratio $R(E_T)$ are unchanged below the knee of the
distribution (see Fig.~1). Beyond it, the $J/\psi$ suppression is increased as a result of the
fluctuation.

We turn next to the numerical results. In ref. \cite{2r} we used for the two parameters of the
model $\sigma_{abs} = 6.7$~mb and $\sigma_{co} = 0.6$~mb. In this case, the computed $J/\psi$
suppression at $E_T \sim 100$~GeV is somewhat too small \cite{2r}. Clearly, we can increase it by
increasing the value of $\sigma_{co}$. However, we then increase the value of the suppression for
peripheral collisions. This, in turn, can lead to some conflict with the SU data (see
\cite{3r} for a discussion on this point). However, recent data 
\cite{10rb} on the $J/\psi$
cross-section in $pA$ collisions, point to a smaller value of $\sigma_{abs}$ - of 4 to 5 mb. With
this value of $\sigma_{abs}$, we can increase $\sigma_{co}$ from 0.6 mb up 
to 1.0 mb without decreasing
the $J/\psi$ suppression for peripheral collisions. In Fig.~2 we present the result of our
calculation using $\sigma_{abs} = 4.5$~mb and $\sigma_{co} = 1$~mb. We see that the 
main features of the data are reproduced. In particular our curve shows a
slight change of curvature at $E_T \sim$~100~GeV, which is entirely due to the effect of
fluctuations - and is seen in the 1998 NA50 data \cite{1r}. 
The physical origin of this change in the slope of $E_T$ is the
following: when approaching the knee of the $E_T$ distribution from
below, the number of participants approaches $2A$ and changes slowly. The
latter is also true for the multiplicity of comovers. Beyond the
knee, the multiplicity increases faster due to the fluctuations and
produces a faster decrease of $R(E_T)$ (see Fig. 1).
\par

We want to stress that the shape of our curve in the lower half of the $E_T$ region (where the
ratio $R(E_T)$ changes rather fast with $E_T$) is sensitive to the relation between $E_T$
and impact parameter. We see from eq.
(\ref{5e}) that this relation depends on the size of the contribution of the intra-nuclear cascade
(parameter $k$). As mentioned above, the value $k = 1/4000$ used here was obtained in \cite{2r}
from the best fit of the correlation between $E_T$ and the energy $E_{ZDC}$ of the zero degree
calorimeter. However, since 
we do not have a totally reliable
expression for the latter,
there is an
uncertainty in the value of $k$. In order
to illustrate its effect on $R(E_T)$, we show in Fig.~2 (dashed line) the result with $k =
1/2000$, i.e. doubling the (comparatively small) contribution of the
intra-nuclear cascade. The effect is concentrated in the lower half of the $E_T$ range. This
uncertainty would not be present if the $J/\psi$ suppression were given as a function of either
$E_T$ or charged multiplicity at mid-rapidities. 
\par

The large $E_T$ structure seen by the NA50 collaboration can also be explained assuming a
deconfining phase transition \cite{11r}. At the energies of the 
Relativistic Heavy Ion Collider (RHIC) at Brookhaven,
it will be possible to determine
which of these two mechanisms is the correct one. 
Indeed, the transverse energy (or the
corresponding energy density) where the structure has been seen by NA50, will be reached at
RHIC well below the knee of the $E_T$ distribution, and, if our interpretation is correct, no
structure will be present. It will, however, appear at higher values of $E_T$ - when reaching the
knee of the $E_T$ distribution at $\sqrt{s} = 200$~GeV. \\

\centerline{ACKNOWLEGMENTS} 

It is a pleasure to thank N. Armesto,
C. Pajares,
C. Gerschel,
C. A. Salgado
and Y. M. Shabelski 
for interesting discussions. We also thank B. Chaurand and M. Gonin 
for providing numerical tables of the NA50 data.
E. G. F. thanks Ministerio de Educacion y 
Cultura of Spain for financial support.

\newpage
\centerline{FIGURE CAPTIONS}
\vspace{1cm}

\noi {\bf FIG. 1.}  The ratio $F(E_T)$ in eqs. (\ref{7e}), (\ref{8e}). \\

\noi {\bf FIG. 2.} The ratio $R(E_T)$ of $J/\psi$ over DY cross-sections, obtained with
$\sigma_{abs} = 4.5$~mb and $\sigma_{co} = 1$~mb, compared to the NA50 data \cite{1r}
\cite{12r}. The full curve corresponds to $k = 1/4000$ \cite{2r} in eq. (\ref{5e}). The dashed
curve is obtained with $k = 1/2000$ (see main text).
The black points correspond to 1996 Pb-Pb data, the black squares correspond to
1998 Pb-Pb data, the white points to 1996 analysis with minimum bias
and the white squares to 1998 analysis with minimum bias.

\newpage


\hspace{-1.2cm}\epsfig{file=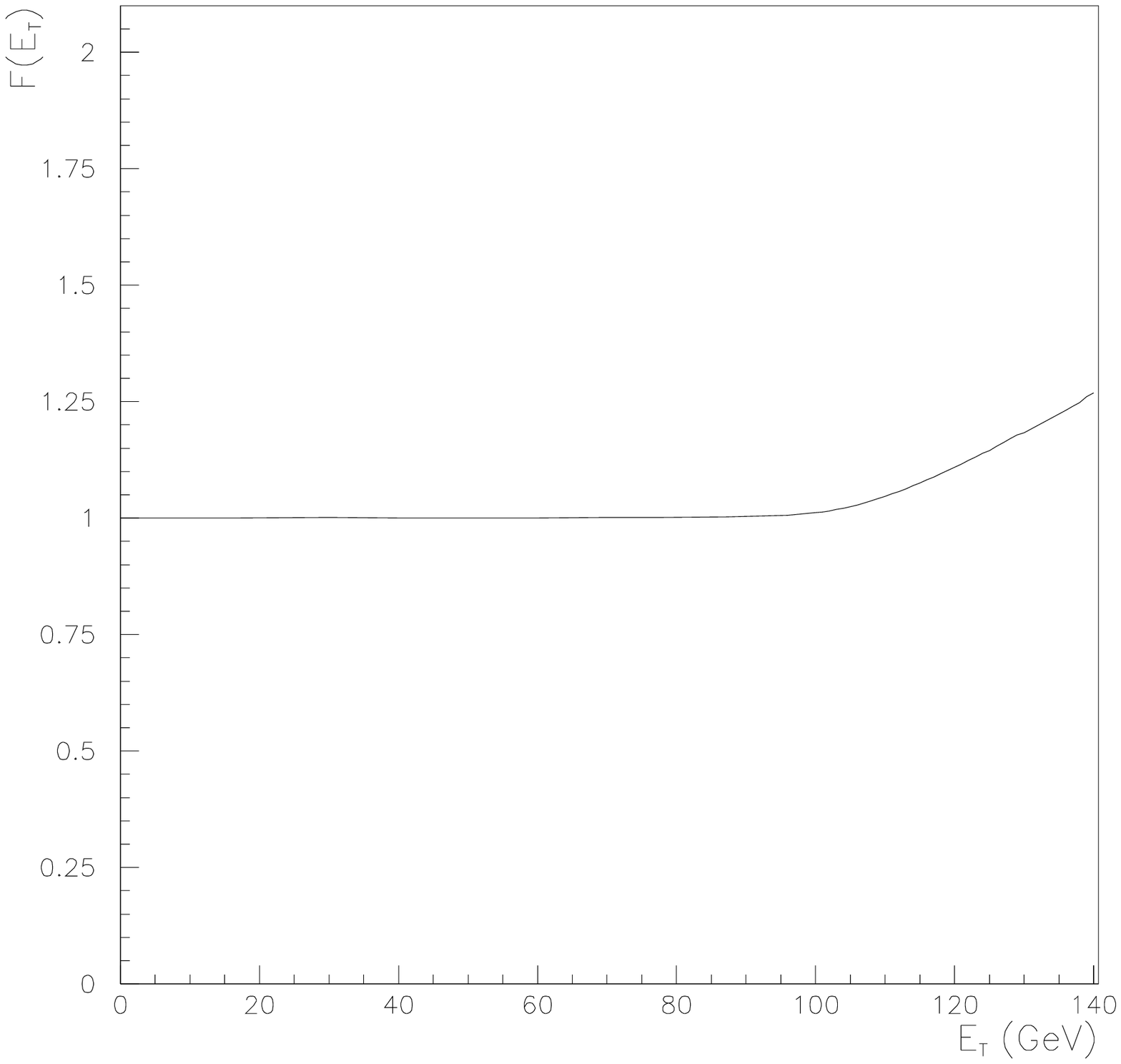,width=16.cm}

\vspace{1cm}
\centerline{1 Capella}

\newpage


\hspace{-1.2cm}\epsfig{file=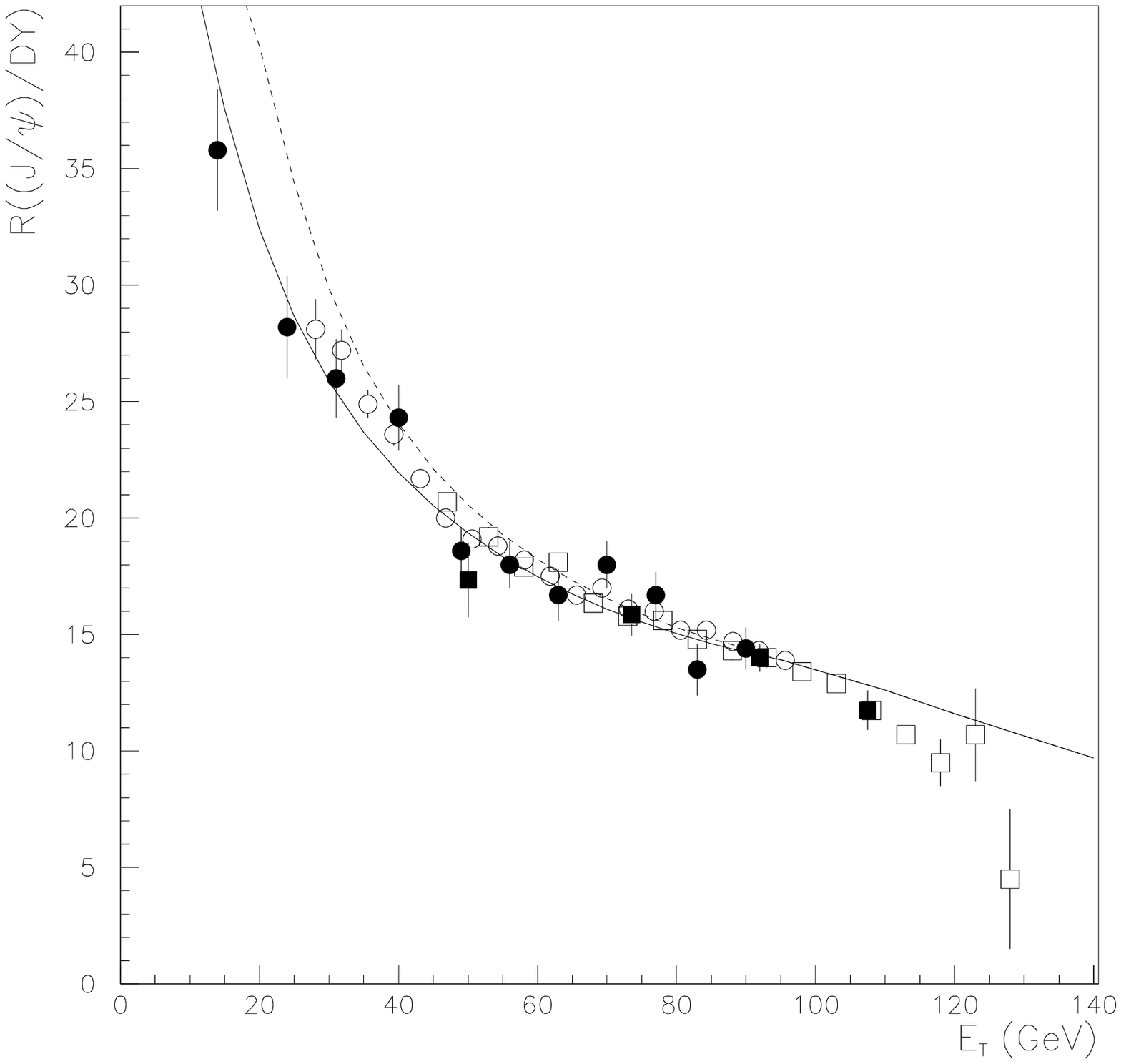,width=16.cm}

\vspace{1cm}
\centerline{2 Capella}

%
%

\end{document}